# Electrically Driven Hot-carrier Generation and Above-threshold Light Emission in Plasmonic Tunnel Junctions

*Longji Cui[1,2,3], Yunxuan Zhu[1], Mahdiyeh Abbasi[4], Arash Ahmadivand[4], Burak Gerislioglu[1], Peter Nordlander[1,4,5] and Douglas Natelson[\*1,4,5]*

[1]Department of Physics and Astronomy and Smalley-Curl Institute, Rice University, Houston, TX 77005

[2]Department of Mechanical Engineering, University of Colorado, Boulder, CO 80309

[3]Materials Science and Engineering Program, University of Colorado, Boulder, CO 80309

[4]Department of Electrical and Computer Engineering, Rice University, Houston, TX 77005

[5]Department of Materials Science and Nanoengineering, Rice University, Houston, TX 77005

 * corresponding author: Douglas Natelson (natelson@rice.edu)

**ABSTRACT**

Above-threshold light emission from plasmonic tunnel junctions, when emitted photons have energies significantly higher than the energy scale of the incident electrons, has attracted much recent interest in nano-optics, while the underlying physical mechanism remains elusive. We examine above-threshold light emission in electromigrated tunnel junctions. Our measurements over a large ensemble of devices demonstrate a giant material dependence of photon yield (emitted photons per incident electrons), as large as four orders of magnitude. This dramatic effect cannot be explained only by the radiative field enhancement effect due to the localized plasmons in the tunneling gap. Emission is well described by a Boltzmann spectrum with an effective temperature exceeding 2000 K, coupled to a plasmon-modified photonic density of states. The effective temperature is approximately linear in the applied bias, consistent with a suggested theoretical model in which hot carriers are generated by non-radiative decay of electrically excited localized plasmons. Electrically driven hot-carrier generation and the associated non-traditional light emission could open new possibilities for active photochemistry, optoelectronics and quantum optics.

**Introduction**

Localized surface plasmons (LSPs) in metal nanostructures are of great current interest for their role in generating non-equilibrium hot carriers for photochemistry [1-3], photodetection [4, 5], photoluminescence [6] and photovoltaics [7, 8]. LSPs and plasmon-induced hot-carrier dynamics can be driven either by optical illumination or electrically via inelastic tunneling. In electrically driven tunnel junctions, plasmon enhanced light emission has been found promising for a variety of technologies that require efficient optoelectronic integration and conversion at the nanoscale [9-16]. Radiative decay of the inelastically excited LSPs in the tunneling gap has been recognized as a dominant light emission mechanism [17], leading to broadband emission at photon energies $\hbar\omega$ less than the "single-electron" energy scale corresponding to the applied voltage (i.e., $\hbar\omega \leqslant eV$) [9, 13, 16, 18-21]. Numerous recent efforts [9, 13, 16, 18] have focused on optimizing plasmonic excitations of metallic nanostructures to improve the light emission efficiency. Interesting, a number of pioneering works [22-27] performed by scanning tunneling microscopy and nanofabricated planar tunnel junctions have reported the observation of above-threshold light emission, where photons emitted from tunnel junctions have energies extending to $2eV$ or even $3eV$, in contradiction with a simple single-electron picture of electrically driven plasmonic excitation and decay.

In contrast to the below-threshold light emission, above-threshold light emission ($\hbar\omega > eV$) requires multi-electron processes. One possible mechanism is based on blackbody thermal radiation of the hot-electron gas formed in the drain electrode by electrons that elastically tunnel through the junction [22, 25]. In this physical picture, tunneling electrons can thermalize rapidly via inelastic electron-electron scattering, faster than their energy can be coupled to the lattice via electron-phonon scattering. The result is a hot electron gas on top of a background of cold carriers, with the hot electrons distribution characterized by an effective temperature set by the dissipated electrical power (Joule heating) and electronic transport of heat. That effective temperature can be much higher than the equilibrium temperature of the lattice. The thermal emission spectrum of this hot electron gas reflects the plasmon-modified photonic local density of states rather than the free-space density of states. Other multi-electron mechanisms[23, 26-29] have also been proposed, under which the coherent interaction between electrons, either through Auger-like processes or mediated by plasmonic excitations, could facilitate inelastic tunneling electrons gaining excess energy above *eV* and subsequently lead to above-threshold light emission via radiative decay of excited LSPs. Despite these mechanisms, it remains a great challenge to experimentally identify the physical origin of the above-threshold light emission.

To investigate this problem, in this work we perform experimental studies on the light emission in tunnel junction devices made of materials chosen to have differing plasmonic properties, and under various electrical driving conditions. A surprisingly strong material-dependent photon yield is observed, as large as $10^4$-fold, which cannot by explained solely by the well-established material-dependent radiative field enhancement (proportional to the photon density of states) due to the LSPs confined in the subnanometer-sized tunneling gap. In pure Au junctions, nearly all of the emitted light can be above threshold. We confirm that the spectrum of the radiation is Boltzmann distributed coupled with the junction-specific plasmon-modified photon density of states, showing that there are hot carriers described by an effective temperature much higher than the lattice temperature. In contrast to prior studies, when looking at the ensemble of junctions we find that this effective temperature is set by the bias voltage rather than dissipated electrical power. We propose a microscopic theoretical model based on hot carriers generated by non-radiative decay of inelastically excited plasmons and find it to be consistent with our experimental results and statistical analysis over a large ensemble of tunnel junction devices.

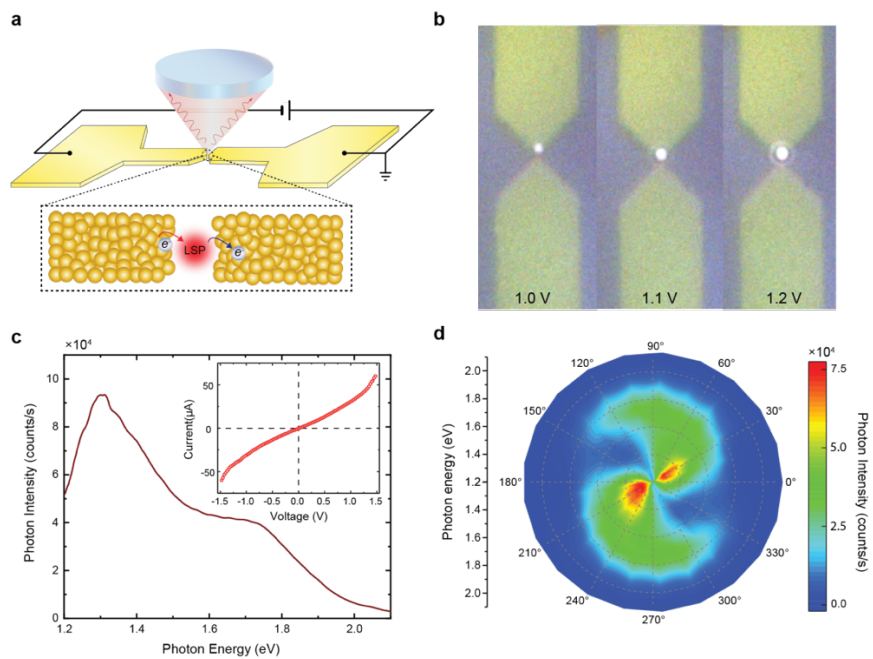

**Figure 1**. Experimental strategy for measuring light emission in electrically driven tunnel junctions. **a**) Schematics of the experimental setup capable of simultaneous electrical transport and optical spectroscopy measurements. LSP denotes the localizes surface plasmons excited by the inelastic tunneling electrons. **b**) Wide-field CCD imaging of an Au light-emitting tunnel junction operating under bias from 1.0 to 1.2 $V$. A weak white light source is applied to illuminate the device structure. **c**) Measured light emission spectrum of the Au tunnel junction at 1.0 $V$. Inset shows the d.c. *I-V* characteristics of the junction. **d**) Polarization-dependent spectral light emission contour plot for the operating tunnel junction in d) at 1.0 $V$. 0° and 90° denotes the polarization along and across the tunnel junction, respectively.

**Results and discussion**

Our experimental approach for measuring light emission in plasmonic tunnel junctions is illustrated in Figure. 1A. We fabricated samples consisting of arrays of nanowires made of several different metallic materials (Au, Au/Cr, $Au_{0.6}Pd_{0.4}$/Cr, and Pd/Cr) (See Supplementary Information for the nanofabrication steps). These materials were chosen deliberately to test the role of plasmons in the emission process, as these metals range respectively from good to poor plasmonic properties in the red part of the visible spectrum. Numerical simulation is conducted to optimize the geometry of the nanowires (600-nm long, 100-nm wide and 18-nm thick) to obtain maximal plasmonic response. The ultrathin adhesion layer of Cr (~1 nm) functions as a damping medium [30, 31] to attenuate plasmonic resonances relative to the pure Au case. Au/Pd and Pd have poor plasmonic properties in this spectral range due to higher resistivity and interband transitions. To create a subnanometer-sized tunneling gap in the nanowire to form a tunnel junction, an electromigration break junction approach[32] was employed (see Supplementary Information). In performing light emission measurements, we applied a voltage bias $V$ to drive electron tunneling through the junction and measured the resultant electrical current and light emission simultaneously. To maintain high stability and cleanliness of the tunnel junctions during the measurements, our experiments were performed with the substrate temperature at 5 K in high vacuum environment. Emitted photons from the tunnel junction were collected through free-space optics and imaged on a CCD camera/spectrometer.

As shown in Fig. 1b, the wide-field light emitting images of the tunnel junction operating under increasing voltage bias $V$ from 1 to 1.2 $V$ clearly demonstrate a voltage-tunable, bright, nanoscale light source (diffraction limited). Figure 1c shows a representative light emission spectrum recorded for a Au tunnel junction device (zero-bias *d.c.* conductance $G = 0.25\ G_0$, where $G_0 = 2e^2/h$ is the conductance quantum). Note that *all* light recorded by the spectrometer are above-threshold photons ($\hbar\omega > 1\ eV$). Furthermore, we performed polarization-dependent measurements of the light emission to examine the plasmonic modes of the tunnel junction. As shown in Fig. 1d, the polarization-spectral contour plot reveals the mode structure of the LSP resonances excited by the inelastic tunneling electrons, with peaks in the photon emission at around 1.3 $eV$ and 1.7 $eV$ for the studied junction (see also in Fig. 1c). Past studies [33] on the plasmonic properties of such tunnel junction devices have shown that both the dipolar "tip" plasmons and the transverse plasmons originating from the nanowire contribute to the plasmonic modes. The hybridization of these and higher order multipolar modes, due to the broken symmetry of the tunnel junction geometry, creates the LSPs confined in the tunneling gap.

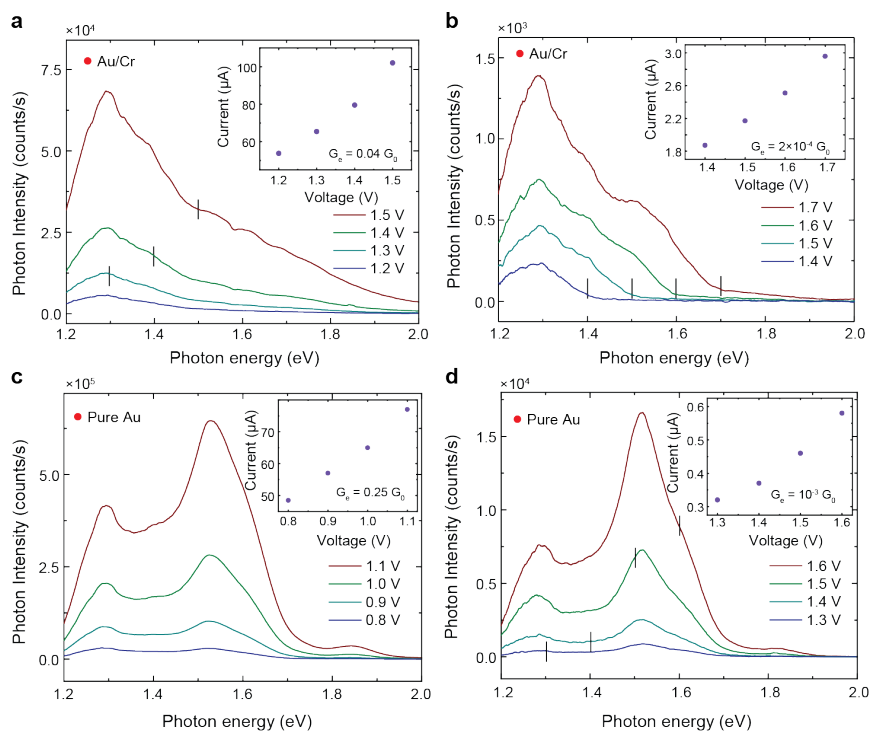

**Figure 2**. Measured above-threshold light emission in different current regimes of pure Au and Au/Cr tunnel junction devices. **a**) Spectral light emission intensity for an Au/Cr tunnel junction in high-current regime (~50–100 µA). The short black lines mark the energy threshold corresponding to the applied voltage (*eV*). **b**) Measured light emission spectra for the same tunnel junction in **a**), but the junction is further electromigrated to form a wider tunnel gap (with lower zero-bias conductance) carrying much lower current (~2 µA). The light emission spectra exhibit clear cut-off at the threshold energy, resembling below-threshold light emission. **c**) Same as **a**), but for a pure Au tunnel junction in high-current regime. *All* emitted photons have energies above the threshold. **d**) The same pure Au junction is further electromigrated to form a low-current carrying device. A sizable portion of the light emission spectra still shows above-threshold photons. Insets in **a**) to **d**) show the corresponding conductance measurements of each junction in different current regimes, and *Ge* is the zero-bias *d.c.* conductance in unit of the quantum conductance ($G_0 = 2e^2/h = 1/(12.9\ k\Omega)$).

We further examined in detail the above-threshold light emission phenomena in tunnel junctions under different driving voltages and tunneling current. As shown in Fig. 2, we performed light emission measurement on the same tunnel junctions in high- and low-current regimes. To elaborate, as described above, controlled electromigration was applied to break the nanowire to form a tunneling gap. After the experiment was done, the same junction was electromigrated further (see Supplementary Information for details), by applying a larger voltage to form a larger separation tunnel gap which is indicated by a much smaller zero-bias conductance compared to the tunnel junction before second electromigration. The tunnel junctions after the first and second electromigration step were found to allow a high (~100 µA) and low (below 10 µA) tunneling current, respectively.

Comparison of the measurement results for Au/Cr and pure Au junctions reveals important insights. In Au/Cr junctions, light emission in high-current regime (Figure. 2a) contains a substantial above-threshold portion, whereas the same device in low-current regime (Figure. 2b) exhibits a clear energy cut-off at $eV$ in the emission spectrum, reminiscent of the previous studies of below-threshold light emission[9, 13, 18]. In contrast, in pure Au junctions, high-current regime (Fig. 2c) shows that all emitted photons have energies exceeding the $eV$ threshold. Moreover, in low-current regime (Fig. 2d), above-threshold photons are still observable (different from that in Fig. 2b for the same current regime), even though the tunneling current in this junction is much smaller than the Au/Cr case. Note here that since device-to-device variation in the obtained conductance of tunnel junctions is significant, it is most meaningful to compare ensembles of devices by photon yield under the same applied voltage, that is, to calculate the total emitted photons per tunneling electrons. We define photon yield $Q$ as $U / I$, where $U$ is the total photon counts and $I$ is the tunneling current. Interestingly, the enhancement factor $Q_{high} / Q_{low}$ between the high- and low-current regime is found to be always larger than 1 (see more results in Supplementary Information from ~30 devices), indicating a non-linear relation between the photon emission and tunneling current.

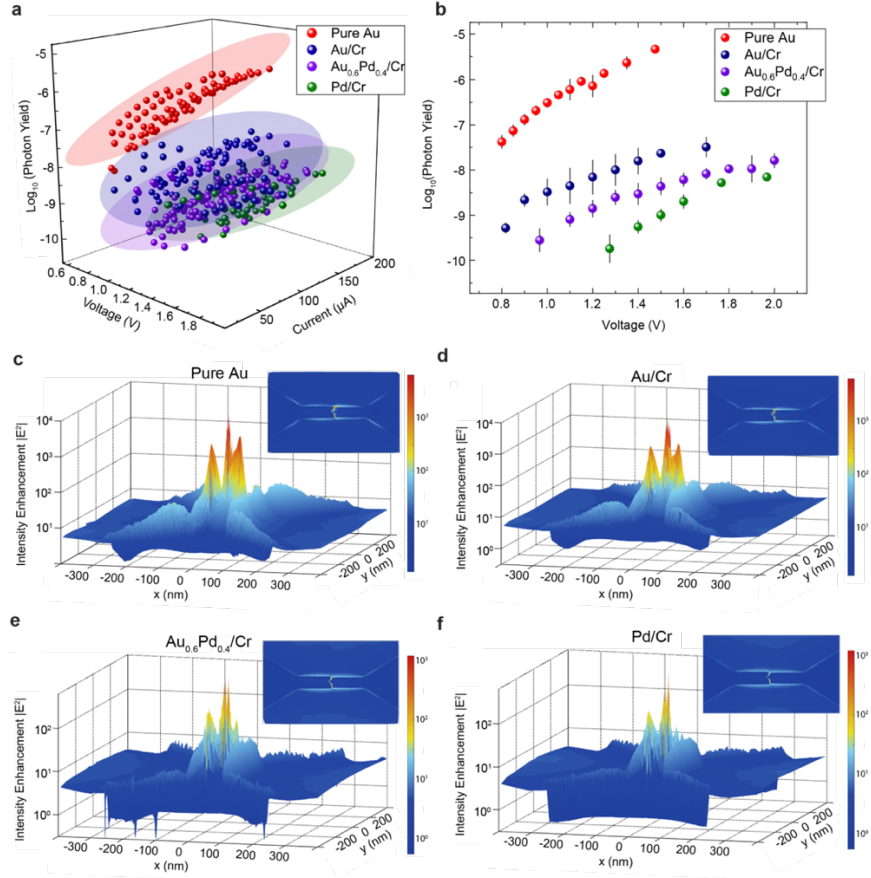

**Figure 3**. Material- and voltage-dependence of above-threshold light emission and simulations for the radiative field enhancement due to the localized plasmons in the tunneling gap. **a**) Measured photon yield (plotted on logarithmic scale) for ~100 tunnel junction devices of different plasmonic materials vs. applied voltage and tunneling current. The ellipsoids correspond to a 95% confidence interval fit to the experimental data. **b**) Measured photon yield vs. applied voltage for ensemble of selected devices of different materials with fixed tunneling current of ~100 µA. Error bars are the standard deviation over the measured junctions for each material at each bias. **c**) Finite-element simulation results of plasmon-induced electric field intensity enhancement (proportional to the photonic density of states in the tunneling gap) at 785 nm (corresponds to the peak wavelength of the observed light emission) for the Au junction. The insets show the top-view of the 3D plots, indicating the geometry of the simulated junction and the 2D intensity. **d**) to **f**) Same as **c**), but for junctions made of other materials. The plasmon-induced enhancement is seen to be smaller than that in **c**).

To understand the observed material-dependent characteristics of above-threshold light emission and the photon yield in Fig. 2, one may simply invoke the material dependence of the strength of plasmonic resonances in different materials. However, given the device-to-device variability of electromigrated tunnel junctions, it is necessary to investigate a large number of tunnel junctions and perform a statistical analysis to reach reliable conclusions. We then conducted systematic measurements on over 100 devices made of different materials under a series of voltages and tunneling currents. Figure 3a summarizes the measurement results by plotting photon yield (on logarithmic scale) as a function of the applied voltage and the corresponding tunneling current. It is clearly seen that pure Au junctions feature the highest photon yield, followed by Au/Cr, $Au_{0.6}Pd_{0.4}$/Cr, and Pd/Cr with the lowest yield.

We can proceed further and select a set of junctions with roughly similar tunneling current (~100 μA, at which level above-threshold light emission was observed for all materials). The purpose of such selection is to examine the relationship between photon yield and applied voltage, excluding the effect of tunneling current. As shown below in our theoretical model, this relationship is critical to reveal the physical origin of above-threshold photon emission. As shown in Fig. 3b, we found that under the same applied voltage and current, the photon yield of pure Au junctions are around 2 to 3 orders of magnitude higher than Au/Cr and $Au_{0.6}Pd_{0.4}$/Cr junctions, and nearly 4 orders of magnitude higher than Pd/Cr junctions. While one may expect a larger photon yield from a good plasmonic material such as pure Au compared to a well-known poor plasmonic material such as Pd and Pd/Cr (with Cr further attenuating the plasmonic response of Pd), the observation of orders-of-magnitude material-dependent discrepancy is surprising.

To show that, plasmon-induced radiative field enhancement (a direct indicator of the strength of the plasmonic resonances) can be inferred directly from the calculation of the electrical field intensity within the tunneling gap (see Supplementary Information for details of our finite-element modelling and more results). While such classical finite element models omit realistic electronic structure and quantum mechanical tunneling effects, these calculations do show overall trends in magnitudes of such optical antenna effects. As shown in Fig. 3c to 3f, we found that the localized plasmons in pure Au junctions indeed lead to highest radiative field enhancement among the studied materials, but only differ from the worst plasmonic material (Pd/Cr) by a factor of ~20, dramatically smaller than the measured photon yield difference in light emission. These observations, combined with numerical simulations, strongly suggest that above-threshold light emission does not originate solely from a simple plasmonic enhancement effect due to LSPs excited in the tunneling gap.

In analogy with prior efforts [25, 26], we performed a normalization analysis on the measured light emission spectra under different voltages by dividing with reference to the spectrum obtained at the highest voltage. The normalization separates the contributions of voltage-independent plasmonic resonances of the tunnel junction, which depends only on junction geometry and the material type, from the voltage-dependent component of observed light emission. Representative measurement results of light emission from a typical Au tunnel junction at 0.8 V, 0.85 V, 0.9 V, and 0.95V are shown in Fig. 4a. After spectral normalization, the reduced emission spectra are plotted in Fig. 4b, which demonstrates that the normalized spectra intensity, plotted on logarithmic scale, decays linearly with the photon energy.

This energy dependence of light emission can then be described phenomenologically by a Boltzmann statistics factor $e^{-\hbar\omega/k_B T_{eff}}$, where $T_{eff}$ is an effective temperature of the hot carriers, out of equilibrium with the lattice and the background of cold electrons. Proper caution is needed to introduce the concept of effective temperature in any driven, non-equilibrium system. In this case, the clear linearity of Fig. 4b shows that an effective Boltzmann factor describes the hot carrier distribution, as in past studies[34-36]. When the typical time interval between tunneling events in a tunneling gap (on the order of 10s of fs for large currents) is much smaller than the relaxation time of the hot carriers (~ hundreds of fs to ps), hot carriers generated by the electrically excited plasmons will undergo many scattering events, forming a non-thermal steady-state distribution, before thermalizing into the lattice phonons[37]. The steady-state distribution can then be parameterized using an effective temperature.

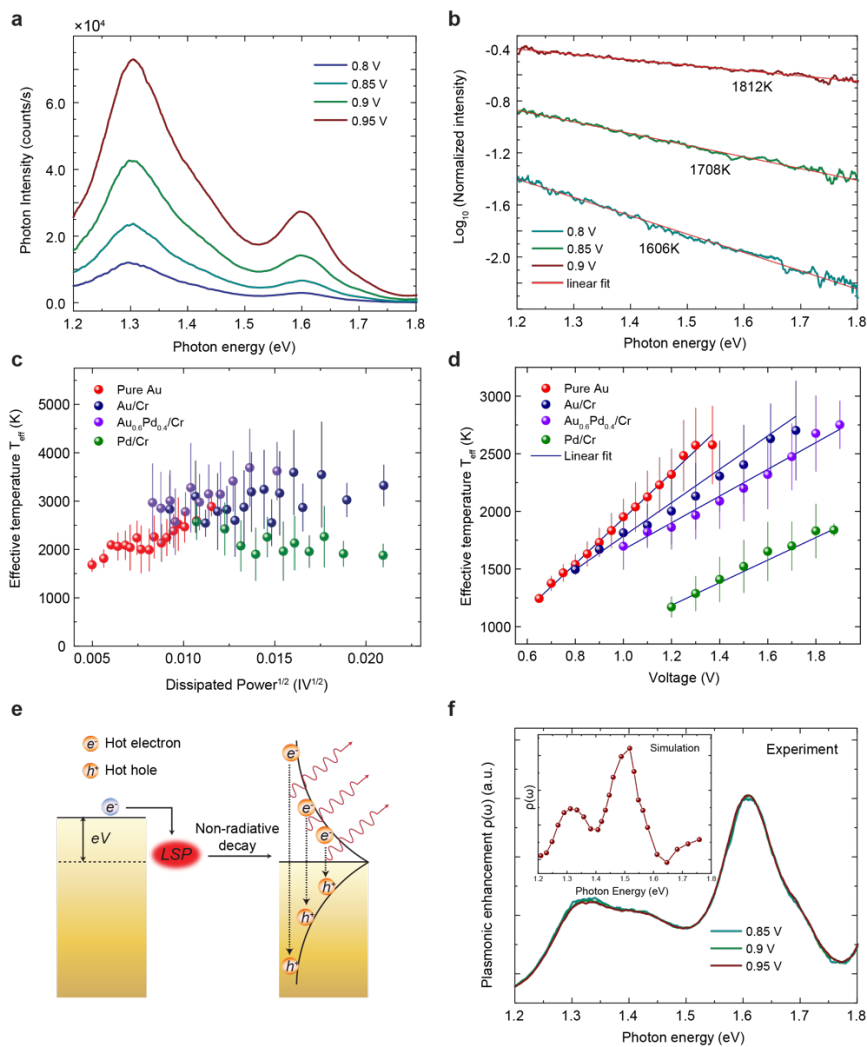

**Figure 4**. Analysis and theoretical model of above-threshold light emission in tunnel junctions. **a)** Measured light emission spectra from an Au tunnel junction under different voltage. **b)** Normalization analysis of the spectra in **a)**, by dividing the measured spectrum at 0.8V, 0.85V, 0.9V with reference to the spectrum at 0.95V. The linear decay of the normalized spectra (on logarithmic scale) is fitted to a Boltzmann energy distribution $e^{-\hbar\omega/k_B T_{eff}}$ (solid red lines), where $T_{eff}$ is the effective temperature of hot energy carriers (electrons and holes). **c)** Statistical analysis over a large ensemble of measured junctions on the extracted effective temperature of hot carriers as a function of power dissipation ($\sqrt{IV}$) in the junction for different material. It can be seen that $T_{eff}$ is relatively independent of the dissipated power, even within a single material. **d)** Statistical analysis on the extracted effective temperature as a function of applied voltage (*V*). The solid lines are the best linear fit to the measured data. Error bars are the standard deviation over the ensemble of measured junctions for each material at the applied voltage. **e)** Schematics of the theoretical model in which light emission comes from the radiative recombination of hot electrons and holes, which is sustained in a steady-state distribution during the continuous electrical excitations of LSPs in the tunneling junction. **f)** Extracted voltage-independent spectral plasmonic enhancement, $\rho(\omega)$, due to LSPs in the tunnel junction from applying Eq. (2). Inset shows the numerically calculated plasmonic enhancement for a tunnel junction with a similar geometry.

We extracted the effective temperature $T_{eff}$ from the normalized emission spectra in Fig. 4b by fitting with a Boltzmann factor. It can be seen that $T_{eff}$ reaches over 1000 K, much higher than the equilibrium lattice temperature of the tunnel junction (i.e., the substrate temperature at 5 K). It appears that above-threshold emission originates from hot carriers. These findings raise the questions: What is the physical explanation of very high non-equilibrium effective temperature? What are the factors determining the effective temperature in such plasmonic system?

To address these questions, we applied the normalization analysis described above to measured emission spectra from the large ensemble of tunnel junction devices. To identify the relationship between the effective temperature and the applied electrical condition (voltage or tunneling current), we evaluate two candidate families of models, $T_{eff} \propto \sqrt{P} = \sqrt{IV}$ and $T_{eff} \propto V$, where $P$ is the dissipated power (Joule heating) in the tunnel junction. Briefly, the first approach is based on the hypothesis that the effective temperature of the hot electrons in the tunnel junction is determined by the electrical power dissipation and coupling between the charge carriers and lattice phonons[25]. In this "thermal" model the blackbody thermal radiation of the hot-electron gas generates broadband above-threshold photons. The second approach considers the electrically driven generation and relaxation of plasmon-induced hot carriers (see below for the microscopic theoretical model).

The results of statistical analysis across the large ensemble of junctions for the first approach is summarized in Fig. 4c. We find that $T_{eff}$ inferred from the normalized emission is largely uncorrelated with the dissipated electrical power, for each of materials studied here. Instead, as shown in Fig. 4d, a linear relationship between $T_{eff}$ and the applied voltage is clearly indicated. Moreover, materials with less plasmonic loss exhibit higher $T_{eff}$ at a given voltage, providing further evidence that plasmons play a key role in the generation of the hot carriers.

Having established empirical correlations of $T_{eff}$ with $V$ directly from our experimental analysis, we sketch a physical model (see schematics in Fig. 4e and detailed derivation in Supplementary Information) to understand this relationship, as well as the generation of above-threshold photons due to plasmon-induced hot-carrier dynamics. While a realistic quantitative theory is extremely challenging, this toy model attempts to capture the essential physics of electrically driven process of hot carrier generation and relaxation. Optical excitation only induces interactions with bright dipolar plasmons. In contrast, inelastic tunneling electrons can excite any localized plasmon (dark and bright) in the tunnel junction [38]. LSPs excited by inelastic tunneling electrons undergo a non-radiative decay process in which a plasmon energy quantum $\hbar\omega_{LSP}$ is transferred to an individual conduction hot electron-hole pair. The energy distribution of these hot carriers is centered around the Fermi level $\varepsilon_F$, and extends to $\varepsilon_F \pm eV$. If the rate of tunneling events outpaces carrier relaxation, a steady-state hot carrier distribution is sustained, with its specific form depending on the time interval between successive electron tunneling events ($\propto e/I$) and the hot carrier lifetimes (~100s of fs)[6]. The above-threshold light emission originates from the plasmon-enhanced radiative recombination of hot electrons and holes with high energies in the hot-carrier distribution.

In this physical picture, a prediction is that a steady-state effective temperature of the hot carriers is directly correlated with the bias window ($eV$) applied to drive the plasmonic process, rather than the dissipated electrical power. Specifically, we found that

$$k_B T_{eff} \propto \beta eV \tag{1}$$

where $\beta$ is material-dependent parameter correlated with the quality of plasmonic response of a material. Eq (1) is quite consistent with our experimental observation in Fig. 4(d) at sufficiently large voltage, and likewise is consistent with the difference in $T_{eff}$ found between plasmonically active and lossy materials at the same voltage. In fact, the dielectric function of transition metals such as Cr and Pd in this energy range dampens the LSPs, thereby decreasing the generation rate of hot carriers through this inelastic excitation mechanism. In addition, the unfilled $d$-band increases the electronic density of states around $\varepsilon_F$, thereby decreasing the lifetimes of excited carriers. This shifts the steady-state distribution to the lower energies (closer to $\varepsilon_F$), making above-threshold light emission less likely [6].

With the insight from this greatly simplified model, a physical picture of the above-threshold light emission can then be obtained. Inspired by a model of anti-Stokes photoluminescence in plasmonic nanoparticles [34, 39] which also shares a hot-carrier origin, we consider the spectral intensity of the light emission

$$U(\omega) \approx \rho(\omega) I^\alpha \hbar\omega e^{-\frac{\hbar\omega}{k_B T_{eff}}} \tag{2}$$

where $\rho(\omega)$ is the photonic density of states (which gives the radiative field enhancement effect due to LSPs for a given junction), $\alpha$ indicates the non-linear tunneling current-dependence of the above-threshold light emission. The value of $\alpha$, always greater than 1, is obtained from the experimental results (see Supplementary Information for more results). Using Eq. (2), combined with the inferred value of $T_{eff}$ from the normalization analysis, our analysis permits the extraction of $\rho(\omega)$. As shown in Fig. 4f, this spectrum shows good qualitative consistency with the electrical field intensity enhancement as a function of photon energy calculated (inset of Fig. 4f) by the finite-element modelling of the plasmonic resonances for a typical device geometry.

The remarkable dependence of photon yield on the plasmonic properties of the constituent metal, a factor of $10^4$ between pure Au and Pd/Cr devices, shows that plasmons play a larger role in the process than just modifying $\rho(\omega)$. Electrically excited, non-radiatively decaying plasmons additionally play a key role in the generation of hot carriers, inducing photon emissions above the energy threshold of single electrons. Our observations and analysis show that hot-carrier distributions with effective temperatures (above 2000 K) are achievable under modest electrical bias (~1 V) in properly designed plasmonic nanostructures, opening avenues for optimization and possible utility in plasmonic chemistry[19] and optoelectronic applications. Similarly, the large plasmonic enhancement of light emission raises possibilities for non-trivial quantum optical effects, as have been seen in other sub-nanometer plasmonic gap systems[40].

**ASSOCIATED CONTENT**

**Supporting Information.**

The following file is available free of charge on the ACS Publication website. Device nanofabrication; Electromigration break junction protocol; Experimental setup; Evaluation of tunneling current dependence of light emission; Additional normalization analysis results in different materials; Influence of substrate temperature on above-threshold light emission; Finite-element modelling; Theoretical model.

# AUTHOR INFORMATION

**Corresponding Author**

*Email: natelson@rice.edu

**Author Contributions**

D.N and L.C. designed the experiment. L.C. and Y.Z. fabricated the devices, conducted the transport and light emission experiment, and theoretically modelled the data. M.A. performed the finite-element numerical simulation. P.N. analytically modelled the driven hot carrier system. A.A. and B.G. performed additional modelling. L.C., D.N., and P.N. wrote the manuscript with comments and inputs from all authors. All authors have given approval to the final version of the manuscript.

**Notes**

The authors declare no competing financial interest.

**Funding Sources**

L.C. acknowledges funding support from J. Evans Attwell Welch Fellowship and the Smalley Curl Institute at Rice University. D.N. and Y.Z. acknowledge Robert A. Welch Foundation award C-1636. D.N. and M. A. acknowledge support for modelling from NSF ECCS-1704625. P.N. acknowledges Robert A. Welch Foundation award C-1222 and AFOSR award No. FA 9550-15-1-0022.

# ACKNOWLEDGMENT

L.C. acknowledges funding support from J. Evans Attwell Welch Fellowship, the Smalley Curl Institute at Rice University, and support from the University of Colorado Boulder. D.N. and Y.Z.

acknowledge Robert A. Welch Foundation award C-1636. D.N. and M. A. acknowledge support for modelling from NSF ECCS-1704625. P.N. acknowledges Robert A. Welch Foundation award C-1222 and AFOSR award No. FA 9550-15-1-0022.

## ABBREVIATIONS

LSP, localized surface plasmon.

**For Table of Contents Only**

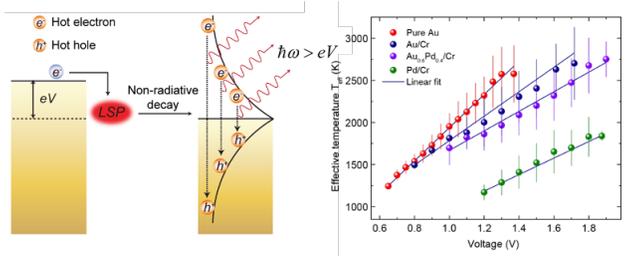

# Supporting Information

# Electrically Driven Hot-carrier Generation and Above-threshold Light Emission in Plasmonic Tunnel Junctions

*Longji Cui[1,2,3], Yunxuan Zhu[1], Mahdiyeh Abbasi[4], Arash Ahmadivand[4], Burak Gerislioglu[1], Peter Nordlander[1,4,5] and Douglas Natelson[*1,4,5]*

[1]Department of Physics and Astronomy and Smalley-Curl Institute, Rice University, Houston, TX 77005

[2]Department of Mechanical Engineering, University of Colorado, Boulder, CO 80309

[3]Materials Science and Engineering Program, University of Colorado, Boulder, CO 80309

[4]Department of Electrical and Computer Engineering, Rice University, Houston, TX 77005

[5]Department of Materials Science and Nanoengineering, Rice University, Houston, TX 77005

 * corresponding author: Douglas Natelson (natelson@rice.edu)

**Supplementary Information Text**

**Device Nanofabrication**

To fabricate the nanowire devices, we start with a 500-μm-thick silicon wafer with a 200-nm-thick thermally grown oxide on top. A shadow mask is applied to evaporate large electrode pads (50-nm-Au/5-nm-Ti) via e-beam evaporation. The nanowire patterns (100 nm wide and 600 nm long) are defined on the spin-coated positive resist (bilayer PMMA 495/950 for pure Au and single layer PMMA 495 for metals with adhesion layers) via e-beam lithography (EBL). 18 nm of metallic layer (Au, $Au_{0.6}Pd_{0.4}$, or Pd) and 1 nm of Cr (when used as the adhesion layer and plasmonic damping medium), are deposited via e-beam evaporation and lift-off to create the nanowires. The devices are cleaned using oxygen plasma and transferred into a vacuum optical cryostat (Montana Instruments, ~$5\times10^{-8}$ Torr) for measurements.

**Electromigration Break Junction Protocol**

The electromigration process is initiated at 80 K by supplying cycles of voltage sweep at a slow rate (10 $mV$/s) into the nanowires using a source meter (Keithley 2400). As shown in Fig. S1a, the electrical current is continuously recorded as a feedback signal for ending the voltage cycle immediately once a sudden drop of current is measured, indicating a small resistance increase (~0.4% higher than the initial resistance) due to the atomic migration in the nanowire. Subsequent cycles of voltage sweep are applied to induce further electromigration. Once the resistance of nanowire reaches a few 100s of Ohms, the substrate temperature is lowered to 5K. The same electromigration procedure is subsequently performed at this temperature until a tunneling gap is formed. The time evolution of the electrical resistance during electromigration before the formation of the tunnel junction (zero-bias *d.c.* resistance $R_0$ larger than 12.9 $k\Omega$, the resistance quantum) is shown in Fig. S1b. We found that such protocol is crucial to achieve a high yield of tunnel junctions with high stability for light emission measurements (as shown in Fig. S1c, stable tunneling current of the electromigrated tunnel junction in a high-current configuration is continuously recorded for over 2000 s).

The low-current regime of a tunnel junction is reached by further electromigration of a high-current configuration. Specifically, the tunnel junction is subject to increased voltage steps (0.5 *V*/step with 5s of dwell time on each step) until a sizable jump in the electrical resistance is observed. The voltage that initiates further electromigration of the tunnel junction to form a more distant tunneling gap is usually larger than the highest voltage applied in measurements for the high-current configuration (after the first step electromigration). This protocol is found to be more reliable than slowly applying a voltage sweep on a high-current configuration in generating a stable tunneling gap without breaking the junction (in a broken junction scenario, $R_0$ is much larger than GΩ level). As shown in Fig. S1d, the tunneling current is recorded in the low-current configuration of the same tunneling junction in Fig. S1c and demonstrates excellent stability.

We note that while electromigrated junctions have comparative stability, by their nature each junction configuration is more time-consuming to produce than is the case in mechanical break junctions of either the flexing or STM techniques. That means that in electromigrated junction experiments it is generally not possible to generate 2D histograms of thousands of *I-V* characteristics.

**Experimental setup**

Following electromigration of the nanowire devices, the optical pathway of light emission from the tunnel junctions collected through a high NA objective (Nikon ×50, NA 0.7) was aligned and calibrated using a home-built Raman spectroscopy setup. When a voltage bias is supplied across the tunnel junction, the resultant electrical current and emitted photons are simultaneously measured via a current pre-amplifier (SRS 570) and an optical spectrometer (Horiba iHR 320/Synapse CCD), respectively. The wide-field images of an operating light-emitting tunnel junction were captured by a CCD camera. All light emission spectral and polarization-dependent measurement results reported in this work were corrected accordingly with the spectral response function, gain setting of the silicon CCD detector and the light collection efficiency of the free space optics.

**Evaluation of Tunneling Current Dependence of Light Emission**

In addition to the results reported in the main text, we have conducted transport and light emission measurements on ~30 tunnel junctions in which each junction is measured in both high- and low-current regimes, to gain better understanding of the nonlinear current-dependent photon yield. As summarized in Fig. S2, the measured photon yield of light emission in a high-current configuration is found to be always larger than that of the low-current configuration when the same voltage is applied. The observed current dependence and photon yield difference cannot be simply explained by the localized surface plasmon induced intensity enhancement effects. It has been demonstrated by past studies[1, 2] that in a sub-nanometer sized gap, the plasmonic near-field effects tend to be saturated or suppressed by quantum tunneling and non-local charge screening.

A power law dependence of emitted intensity on tunneling current, $U(\omega) \propto I^\alpha$, can be expected, with the exponent larger than 1 as shown in Fig. S2. A linear dependence ($\alpha = 1$) would be expected for traditional single-electron plasmon-based tunneling light emission, that is, the number of emitted photons is proportional to current multiplied by the yield of radiatively decaying plasmon excitations. In our modelling (see below) to quantitatively understand the above-threshold light emission, we have assumed values of $\alpha$ is between 1 and 2, suggested by our measured range in Fig. S2. Specifically, $\alpha = 1.2$ was applied to obtain the modelling results in Fig. 4 in the main text.

**Additional Normalization Analysis Results in Different Materials**

The normalization analysis we applied in this work is to enable the comparison and evaluation of measured above-threshold light emission spectra by ruling out the influence from the device-to-device and material-to-material variations in plasmonic properties. To demonstrate the validity of our procedure, additional results for the light emission spectra and the analysis of the electrically driven hot carrier mechanism are plotted in Fig. S3. It can be seen that the linear decay behavior of the normalized spectral intensity (on logarithmic scale) is common to results from tunnel junctions made of all materials, strongly suggesting the universality of such behavior in above-threshold light emission.

A Boltzmann factor $e^{-\hbar\omega/k_B T_{eff}}$, which is used to well approximate the normalized intensity, can be understood as the steady state hot-carrier distribution component over a broadband energy scale below and above $eV$. The slope of the linear fit curves, which are always negative, are found to be increasing with the applied voltage (and also the tunneling current) in all measured devices with above-threshold light emission, indicating that $T_{eff}$ is a function of the electrical transport properties of the tunnel junction. To validate the hot-carrier induced light emission mechanism, we can further extract the value of $T_{eff}$ and reveal the relationship between $T_{eff}$ and different transport quantities (applied voltage or dissipated electrical power) based on the normalized spectral intensity.

Figure S3 also shows that the collapsed plasmonic resonance function (directly related to the optical local density of state of the localized surface plasmons in the tunnel junctions) obtained at different voltages is found to exhibit sharper resonance peaks in pure Au junctions compared to those obtained for other materials. This is expected, since the thin Cr adhesion layer functions as a damping medium that significantly attenuates the plasmonic resonances of a material.

**Influence of Substrate Temperature on Above-threshold Light Emission**

To rule out that physical mechanisms related to the substrate temperature contribute to the observe above-threshold light emission, we conducted several control experiments at different temperatures. Figure S4 shows the light emission spectra collected for a tunnel junction measured at different temperatures (5 K, 30 K and 80 K). It can be seen that the overall line shape of the spectra is not altered, and the normalized spectral intensity curves show the linear decay as expected. The slightly modified emission intensity among different temperatures are most likely attributed to the minute atomic migration of the tunnel junction that is thermally driven, which will correspondingly influence the plasmonic properties of the junction. The finite temperature effect has been considered in previous work[3,4] to understand the observed above-threshold light emission due to the fact that at higher temperature, the Fermi-Dirac distribution of electrons near the energy threshold of $eV$ is broadened by an amount on the order of a few $k_BT$ (~26 $meV$). It can be seen from Fig. S4 that such finite temperature effect is unlikely to explain the measured above-threshold emission in which photons with energies exceeding the energy threshold by around 1 $eV$ are observed. Instead, the observed features on the light emission spectra and the normalized spectral intensity at different temperatures strongly suggest that a plasmonic (non-thermal) origin contributes to the light emission.

**Finite-element Modelling**

We note at the outset: A full quantum treatment of the plasmonic electrodynamics of the nanogap structures, including realistic electronic structure of the metals and a quantum treatment of the interelectrode tunneling, is well beyond the scope of this work. The purpose of this classical modeling is to establish the relevance of standard optical antenna effects to the observed mode structure, and to estimate roughly the material dependence of expected local field enhancements. The finite-element modelling (FEM) of the plasmonic resonances of tunnel junctions is performed using COMSOL™ (Electromagnetic Waves Frequency Domain Physics). The metallic nanowire structure with an asymmetric gap structure (resembling the geometry of a typical tunneling junction studied here) on top of 200-nm-thick $SiO_2$ is calculated to obtain the electrical field intensity enhancement ($|E|^2$) at different wavelengths in the tunneling gap. While detailed quantitative results with 1 nm gap dimensions cannot be fully trusted (as they omit realistic electronic structure and quantum tunneling effects), the qualitative results should be robust.

A linearly polarized plane wave (both in the longitudinal and transverse directions) with an electrical field amplitude of $2.5 \times 10^5$ V/m is applied normally incident on the metallic nanowire to induce the plasmonic excitations. Results from both polarizations are used to calculate the total electrical field intensity enhancement at different wavelengths to reveal the dominant plasmonic modes of the nanowire and the enhanced electromagnetic field is assessed in the tunneling gap region. The wavelength dependent relative permittivity of Au, Cr, $Au_{0.6}Pd_{0.4}$ and Pd are taken from previous work[5-8].

In addition to the calculation shown in Fig. 3d to 3f in the main text, which are for a gap size of 1 nm and at 785 nm, we have also conducted numerical simulations at different wavelength and gap sizes. Figure S5a to d shows the calculation results for plane waves at 955 nm in a tunnel junction with the gap size of 1 nm, which corresponds to the observed typical low energy light emission peak. It can be seen that the calculated plasmonic enhancement increased by approximately a factor of 10-20 from Pd/Cr to Au tunnel junctions, similar to that shown in Fig. 3. Furthermore, the plasmonic mode and enhancement effects in a tunnel junction with gap size of 4 nm was also evaluated. As shown in Figure. S5e to h, the electric field intensity enhancement is found to be smaller in the larger gap scenario as expected for a gap that is beyond the tunneling regime. Similar to the 1 nm gap case, the plasmonic intensity enhancement effects among different materials only differ by one order of magnitude, much smaller than the four orders of magnitude of difference in the photon yield of above-threshold light emission. These numerical calculations provide further details to validate our claim that the observed above-threshold light emission and the giant material dependence in photon yield cannot be simply explained by the plasmonically enhanced optical density of states in the tunneling gap.

**Theoretical Model**

A full, rigorous treatment of the biased plasmonic tunnel junction system is well beyond the scope of the present work, and would require detailed modeling of carrier transport, inelastic electron-electron and electron-phonon processes, and elastic scattering from disorder and surfaces. Instead, here we develop a very simplified theoretical model to illustrate the plausibility of the experimentally extracted phenomenological relationship between the effective temperature of hot carriers and the applied voltage.

As mentioned in the main text, it is generally not true that driven, nonequilibrium systems can be described well by an effective temperature. Empirically, the linear dependence of the log of the normalized intensity on frequency shows that one can define an effective Boltzmann factor. The key idea in this simple model is that the effective temperature $T_{eff}$ can be related to the average energy content per carrier ($\bar{E}$) of the hot carriers in the steady state distribution, which can be written as

$$k_B T_{eff} = \bar{E} = \frac{M_1}{M_0} \tag{1}$$

where $k_B$ is the Boltzmann constant, $M_1$ and $M_0$ corresponds to the total steady state energy content of the hot carriers and the steady state carrier population, respectively. Given the steady state hot-carrier energy distribution ($n_{ss}(\varepsilon)$ per unit energy), the total energy of the hot carrier system can be expressed as

$$M_1 = \int n_{ss}(\varepsilon) \cdot \varepsilon d\varepsilon \tag{2}$$

where $\varepsilon$ denotes the energy of the excited carriers (with $\varepsilon = 0$ at the Fermi level).

For an open system such as the tunnel junctions studied in this work, the energy carriers involved in the plasmonic processes will flow in and out of the system due to ordinary and bias-driven diffusion. To reach steady state, the number of carriers that flow through the tunnel junction during an effective lifetime scale $\tau_0$ can be written as

$$M_0 \propto I\tau_0 \tag{3}$$

where $I$ is the tunneling current. Therefore, according to Eq. (1), $T_{eff}$ can be written as

$$k_B T_{eff} \propto \frac{\int n_{ss}(\varepsilon) \cdot \varepsilon d\varepsilon}{I\tau_0} \tag{4}$$

To obtain the steady state hot-carrier distribution $n_{ss}(\varepsilon)$, knowing the strong role that plasmons play in these devices from the empirically observed giant material dependence, we consider the creation of hot carriers via a plasmon-mediated process. A tunneling electron can inelastically excite a localized surface plasmon that decays non-radiatively into a hot electron-hole pair, generating carriers within the energy interval of width $2eV$ around the Fermi level. The generated hot carriers will undergo relaxation processes via electron-electron and electron-phonon scattering into lower electronic energy states, eventually reaching a thermally equilibrated distribution. For a tunnel junction with high tunneling current (i.e., in high-current regime), the time spacing of successive tunneling electrons (proportional to $e/I$) will be shorter than the time scale of hot-carrier relaxation. Therefore, the hot carriers will form a quasi-steady state in which the carrier distribution is in a dynamic balance during the fast excitation of localized plasmons. As the time spacing of electron tunneling decreases, one would expect that more hot carriers with high energies will be left in the distribution because of the lack of time to relax to lower energy states.

To describe this physical process, rather than keeping track of the detailed electron-electron inelastic scattering which is responsible for the formation of the steady-state hot-carrier distribution, we focus on a very simplified model that keeps the essential feature of the decline with time of the carrier population at high energies. Again, this is *not* claimed to be a realistic treatment of carrier relaxation. At a given energy state $\varepsilon$, the probability as a function of time of a hot carrier created at energy $\varepsilon$ remaining some time *t* later is crudely approximated as[9]

$$P(\varepsilon,t) = exp(-\gamma|\varepsilon + \delta|t) \tag{5}$$

where $\gamma$ is the decay parameter that is related to the electron-electron inelastic scattering rate, and is thus material-dependent. The parameter $\delta$ is inserted to avoid divergences associated with carriers at the Fermi energy living forever.

We define $C_e$ as the probability per unit energy of a hot carrier being created at energy $\varepsilon$ due to decay of an inelastically excited localized plasmon. $C_e$ is material dependent, and would be larger in materials with good plasmonic properties of interest (e.g., Au) than in materials with comparatively poor plasmonic properties.

We model the effect of regular consecutive electronic excitations of plasmons and subsequent creation and relaxation of hot carriers by[9]

$$n_{ss}(\varepsilon, t) = C_e(P(\varepsilon, t) + P(\varepsilon, t + T_e) + P(\varepsilon, t + 2T_e) + P(\varepsilon, t + 3T_e) + \cdots)$$

$$= C_e(e^{-\gamma(|\varepsilon|+\delta)t} + e^{-\gamma(|\varepsilon|+\delta)(t+T_e)} + e^{-\gamma(|\varepsilon|+\delta)(t+2T_e)} + e^{-\gamma(|\varepsilon|+\delta)(t+3T_e)} + \cdots)$$

$$= \frac{C_e e^{-\gamma(|\varepsilon|+\delta)t}}{1 - e^{-\gamma(|\varepsilon|+\delta)T_e}} \quad (6)$$

where $T_e$ is the time spacing between successive hot carrier excitations and is related to tunneling current by $T_e \propto e/I$. Eq. (6) describes the contribution of the decay of all excited carriers before a given time point $t$. The steady state distribution $n_{ss}(\varepsilon)$ can be obtained by performing a time average over one excitation cycle on the instantaneous distribution in Eq. (6), giving

$$n_{ss}(\varepsilon) = \frac{1}{T_e}\int_0^{T_e} n(\varepsilon, t) dt = \frac{C_e}{\gamma(|\varepsilon| + \delta)T_e} \quad (7)$$

Again, in Eq. (6) and (7), we introduce a small damping $\delta$ in the energy term $|\varepsilon|$ to avoid the divergence problem in the integral at $\varepsilon = \varepsilon_F = 0$ if carriers at the Fermi level had infinite lifetimes. Hot carriers produced by decay of plasmons excited inelastically by tunneling electrons are created in the interval $\varepsilon \in (-eV, eV)$, so Eq. (7) should be written as

$$n_{ss}(\varepsilon) = \frac{C_e}{\gamma(|\varepsilon| + \delta)T_e}\big(\Theta(eV) - \Theta(-eV)\big) \quad (8)$$

Inserting Eq. (8) into Eq. (4) and applying the relation $I \propto e/T_e$, the effective temperature of hot carriers can be obtained as

$$k_B T_{eff} \propto \frac{\int_{-eV}^{eV} n_{ss}(\varepsilon) \cdot \varepsilon d\varepsilon}{I\tau_0} \propto \frac{C_e}{\gamma\tau_0}\left(eV - \delta \ln\left(1 + \frac{eV}{\delta}\right)\right) \quad (9)$$

Noticing that $\delta$ is a small damping factor (~250 meV calculated in past studies[9-11], representing the decay rate of carriers near the fermi level), under sufficiently high driving voltage, Eq. (9) can be approximated as

$$k_B T_{eff} \propto \beta eV \quad (10)$$

which is in excellent agreement with the empirical model we have obtained from the normalization and statistical analysis of our measurements over large amount of devices. Moreover, $\beta$ is a material dependent parameter that takes on a higher value for a more plasmonically active material, consistent qualitatively with what we have seen on Fig. 4d in the main text among different materials. As shown in Fig. S6, we plotted the comparison between two hypothetical plasmonic materials. For the sake of simplicity, we assumed different material-dependent parameters. While developing a detailed, realistic theoretical model of this driven, open system will provide deeper insights, this simplified toy model captures essential physics that shows good consistency with our experimental observations: successive inelastic excitation of plasmons that then generate hot carriers in an interval of $eV$ around the Fermi level can lead to a steady state average energy per carrier (a proxy for an effective temperature) that is approximately proportional to $V$.

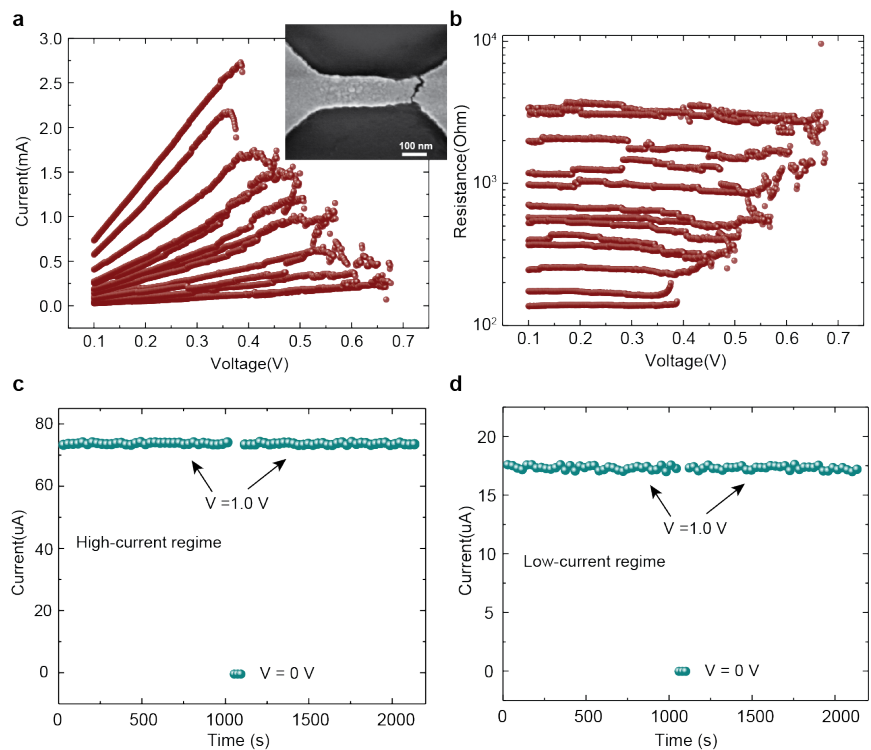

**Figure S1**. Electromigration process to create subnanometer tunneling gaps. a) Measured electrical current (*I*) vs. applied voltage (*V*) during the process of electromigration before the formation of a tunneling gap in the nanowire. The SEM image of the post-electromigration device is shown as the inset. b) Evolution of the electrical resistance (*R*) in the process of electromigration shown in a) until the resistance reaches above the quantized resistance (12.9 kΩ). c) Measurement of the tunneling current stability during an extended period of time. d) Current stability test on the same tunnel junction in c) after further electromigration to obtain a wider tunneling gap.

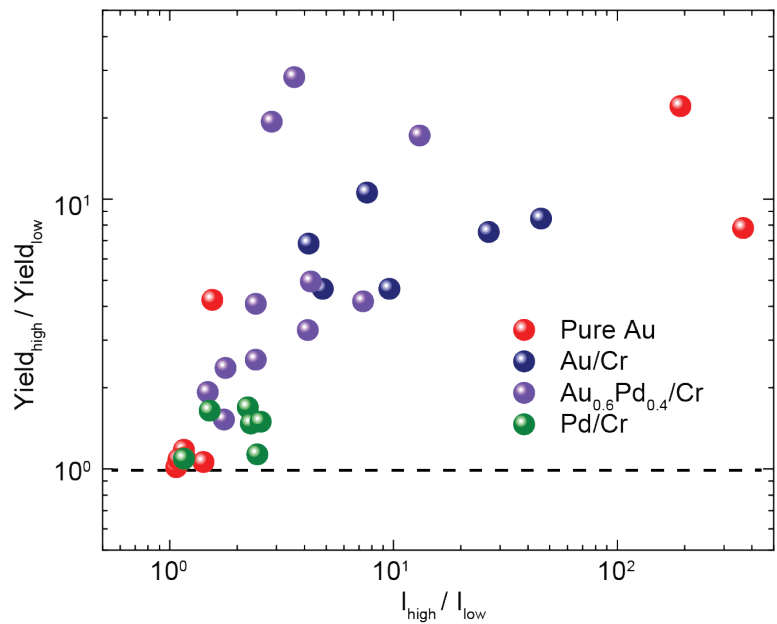

**Figure S2.** Measured enhancement of photon yield in high-current configuration of tunnel junctions. The ratio of the photon yield of high-current configuration to that of the corresponding low-current configuration of the same tunnel junction device (after further electromigration) is plotted as a function of the ratio of the measured electric current in the two configurations.

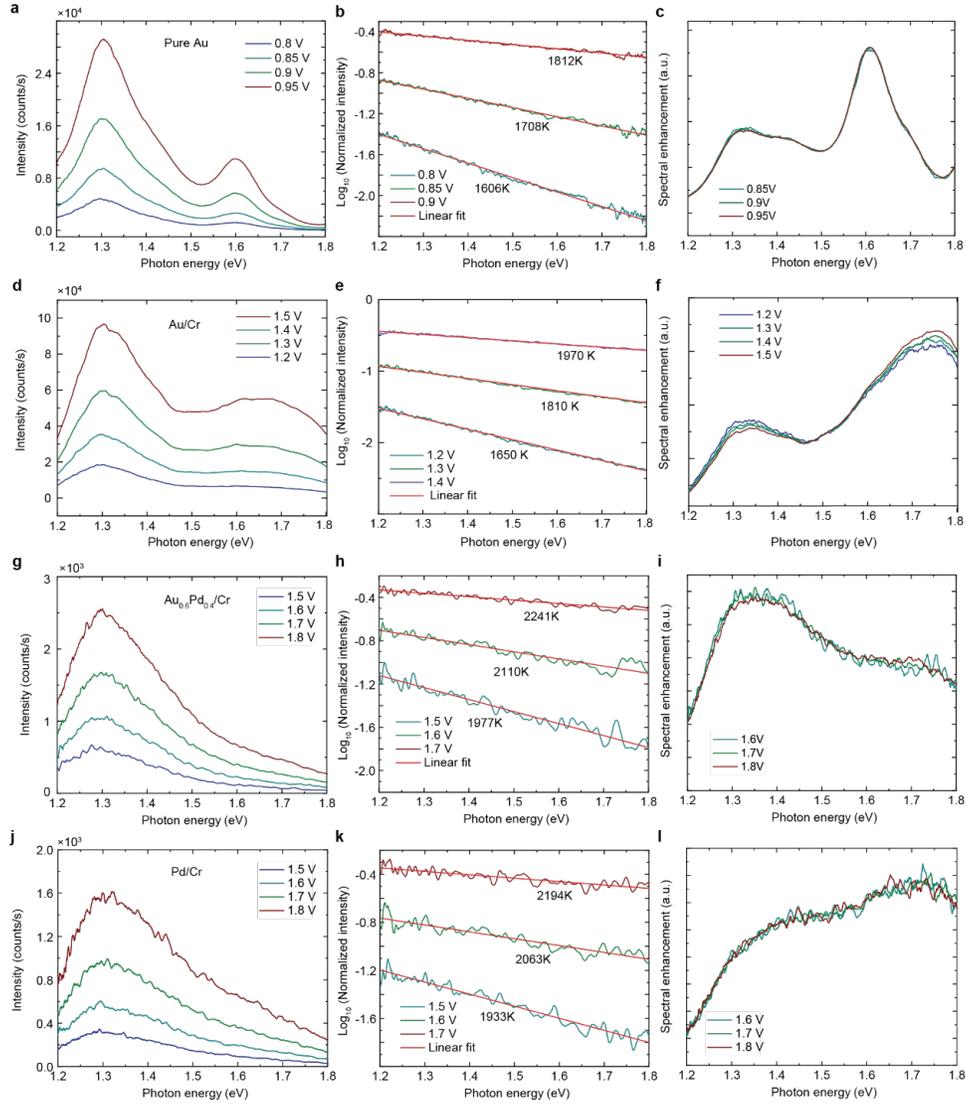

**Figure S3.** Additional results of normalization analysis in different materials. a) to c) show the measured light emission spectra, normalized spectral intensity, and the obtained plasmonic resonance function, respectively, for a pure Au tunnel junction (same as in Fig. 4 in the main text). The extracted effective temperatures of the hot carriers are indicated in b). d) to l) show the results for Au/Cr, $Au_{0.6}Pd_{0.4}$/Cr, and Pd/Cr tunnel junction device, respectively.

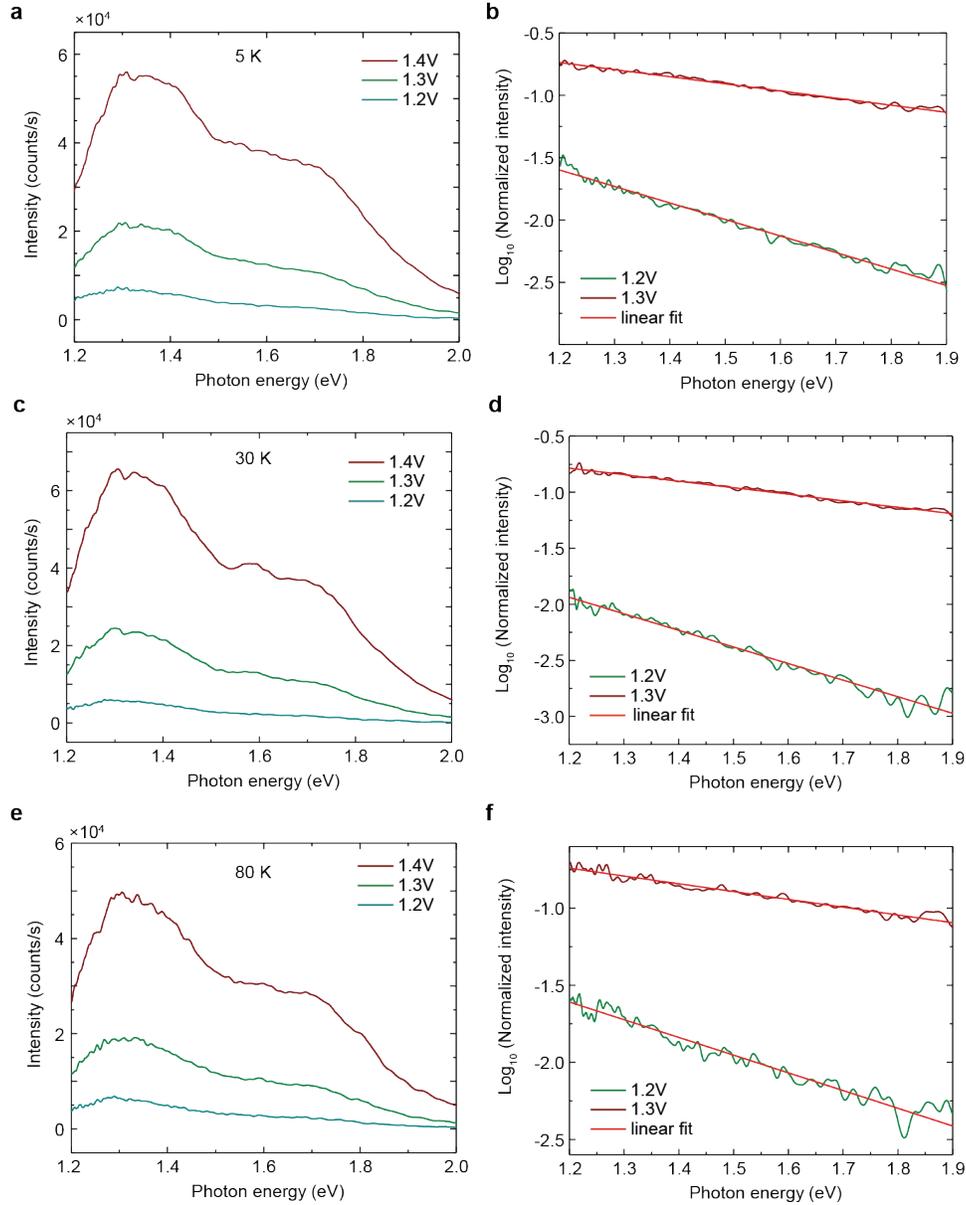

**Figure S4.** Results of control experiments performed on the same tunnel junction devices at different temperatures. a), c) and e) Measured spectral light emission intensity for an Au/Cr tunnel junction at 5 K, 30 K and 80 K, respectively. b), d) and f) Normalized spectral intensity (on logarithmic scale) of the measured spectra in a), c) and e), respectively.

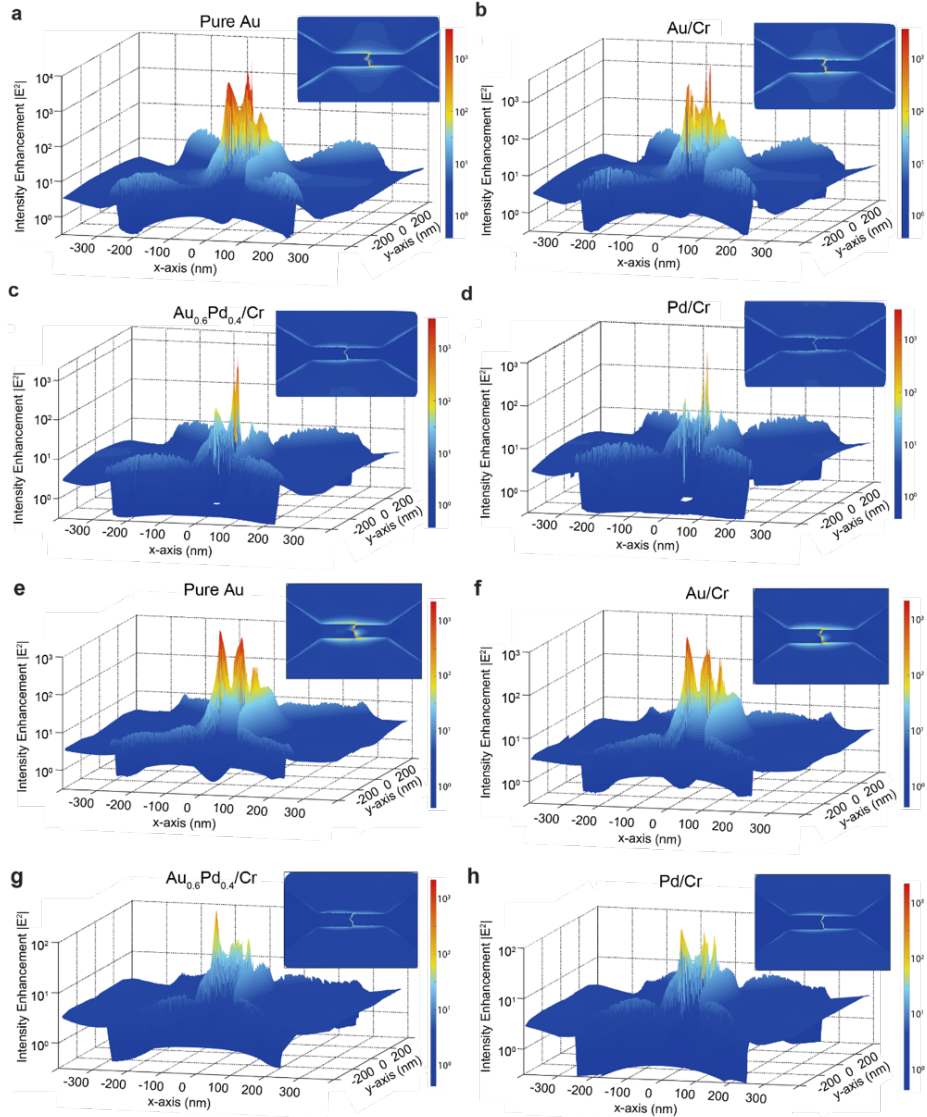

**Figure S5.** Calculated electric field intensity enhancement in the tunneling gap for different wavelength and gap sizes. a to d) show the calculation for 955 nm (~1.3 *eV*, corresponding to the typically observed low energy light emission peak) for different materials with 1 nm tunneling gap. The gap size in this simulation is the same as that in Fig. 3c to 3f. e to h) show the results for 785 nm (corresponding to the high energy light emission peak) for different materials with 4 nm tunneling gap. The insets show the top-view of the 3D plots, indicating the geometry of the simulated junction and the intensity enhancement.

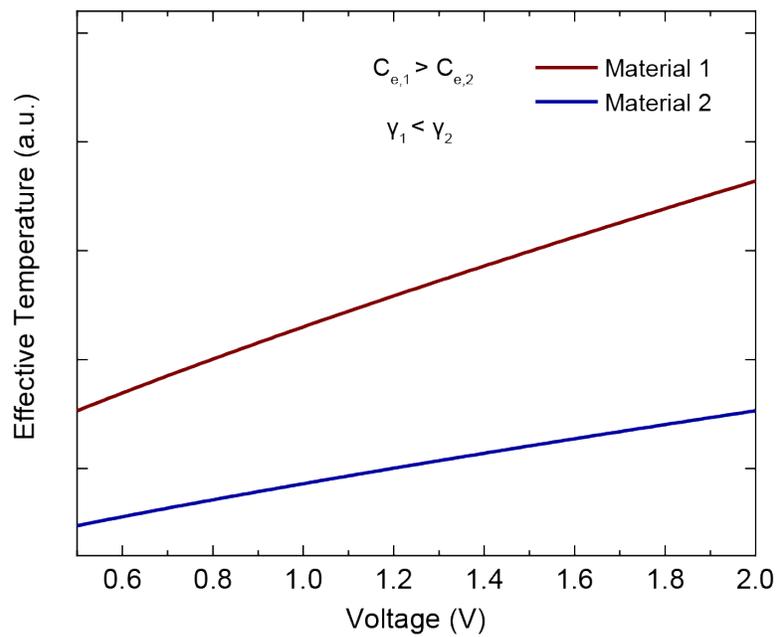

**Figure S6.** Calculated effective temperature as a function of applied voltage according to Eq. (9). $C_e$ and $\gamma$ are material dependent parameters, indicating the strength of plasmonic response in the material. Plasmonically less lossy material (i.e., Material 1) has larger $C_e$ and smaller $\gamma$.